\documentclass[nofootinbib,a4paper,aps,prd,10pt,superscriptaddress,showkeys,twocolumn]{revtex4}

\usepackage{graphicx}
\usepackage{amsfonts}
\usepackage{amssymb} 
\usepackage{amsmath}
\usepackage{hyperref}
\usepackage{natbib}
\usepackage{float}
\usepackage{dcolumn}

\usepackage{bm}

\usepackage{latexsym,color}

\newcommand{\bea}{\begin{eqnarray}}
\newcommand{\eea}{\end{eqnarray}}
\newcommand{\be}{\begin{equation}}
\newcommand{\ee}{\end{equation}}

\begin{document}

\title{Accretion disks properties around regular black hole solutions obtained from non-linear electrodynamics}

\author{Yergali~\surname{Kurmanov}}
\email[]{kurmanov.yergali@kaznu.kz}
\affiliation{National Nanotechnology Laboratory of Open Type,  Almaty 050040, Kazakhstan.}
\affiliation{Al-Farabi Kazakh National University, Al-Farabi av. 71, 050040 Almaty, Kazakhstan.}
\affiliation{International Engineering Technological University, Al-Farabi av. 93G/5, 050060 Almaty, Kazakhstan.}

\author{Kuantay~\surname{Boshkayev}}
\email[]{kuantay@mail.ru}
\affiliation{National Nanotechnology Laboratory of Open Type,  Almaty 050040, Kazakhstan.}
\affiliation{Al-Farabi Kazakh National University, Al-Farabi av. 71, 050040 Almaty, Kazakhstan.}
\affiliation{Institute of Nuclear Physics, Ibragimova, 1, 050032 Almaty, Kazakhstan.}

\author{Talgar~\surname{Konysbayev}}
\email[] {talgar\_777@mail.ru}
\affiliation{National Nanotechnology Laboratory of Open Type,  Almaty 050040, Kazakhstan.}
\affiliation{Al-Farabi Kazakh National University, Al-Farabi av. 71, 050040 Almaty, Kazakhstan.}

\author{Orlando~\surname{Luongo}}
\email[]{orlando.luongo@unicam.it}
\affiliation{Al-Farabi Kazakh National University, Al-Farabi av. 71, 050040 Almaty, Kazakhstan.}
\affiliation{Universit\`a di Camerino, Via Madonna delle Carceri 9, 62032 Camerino, Italy.}
\affiliation{SUNY Polytechnic Institute, 13502 Utica, New York, USA.}
\affiliation{Istituto Nazionale di Fisica Nucleare, Sezione di Perugia, 06123, Perugia,  Italy.}
\affiliation{INAF - Osservatorio Astronomico di Brera, Milano, Italy.}

\author{Nazym~\surname{Saiyp}}
\email[]{nazymsaiyp1@gmail.com}
\affiliation{Al-Farabi Kazakh National University, Al-Farabi av. 71, 050040 Almaty, Kazakhstan.}

\author{Ainur~\surname{Urazalina}}
\email[]{y.a.a.707@mail.ru}
\affiliation{National Nanotechnology Laboratory of Open Type,  Almaty 050040, Kazakhstan.}
\affiliation{Al-Farabi Kazakh National University, Al-Farabi av. 71, 050040 Almaty, Kazakhstan.}
\affiliation{Institute of Nuclear Physics, Ibragimova, 1, 050032 Almaty, Kazakhstan.}

\author{Gulfeiruz~\surname{Ikhsan}}
\email[]{gulfeyruz.ihsan@mail.ru}
\affiliation{National Nanotechnology Laboratory of Open Type,  Almaty 050040, Kazakhstan.}

\author{Gulnara~\surname{Suliyeva}}
\email[]{sulieva.gulnara0899@gmail.com}
\affiliation{National Nanotechnology Laboratory of Open Type,  Almaty 050040, Kazakhstan.}
\affiliation{Al-Farabi Kazakh National University, Al-Farabi av. 71, 050040 Almaty, Kazakhstan.}
\affiliation{Fesenkov Astrophysical Institute, Observatory 23, 050020 Almaty, Kazakhstan.}

\date{\today}

\begin{abstract}
We investigate a family of spherically symmetric, static, charged regular black hole solutions derived within the framework of Einstein-nonlinear electrodynamics. Our study focuses on examining the characteristics of accretion disks in the spacetimes described by the Dymnikova and Fan-Wang solutions. We explore circular geodesics of test particles and calculate various properties, including the radius of the innermost stable circular orbit, radiant energy, temperature, and conversion efficiency of accretion mass into radiation. We employ the Novikov-Thorne-Page thin accretion disk model as a background. By comparing our findings with those obtained in the Schwarzschild black hole case, we reveal significant modifications in the overall spectral properties. Specifically, we observe an increase in the energy emitted from the disk surface, resulting in higher temperatures for the accretion disks under certain values of the free parameters. Consequently, we note an enhanced efficiency of mass conversion into radiation compared to the Schwarzschild spacetime.
\end{abstract}


\maketitle
 
\section{Introduction}
Unlike black holes (BHs), non-singular or regular black holes (RBHs) have been predicted based on specific energy-momentum tensors, introducing additional effects. Recent findings indicate that the violation of the strong energy condition (SEC) is a key feature of RBHs. It's generally observed that in any static region of the event horizon, SEC is violated, resulting in a negative Tolman mass \cite{Zaslavskii}. Conversely, in the non-static case, this property leads to a violation of the dominant energy condition.

The search for RBHs originated from contributions by Sakharov and Gliner \cite{Sakharov, Gliner}, who proposed that introducing a medium with a de Sitter metric instead of a vacuum could potentially circumvent the existence of singularities. This idea was further developed by Dymnikova, Gurevich, and Starobinsky \cite{GlinDym, Gurevich, Starobinskii}, who proposed an intricate interplay between gravitational and electromagnetic fields.

Specifically, non-linear electrodynamics (NLED) can produce fields that interact with extremely strong gravitational sources, particularly in the vicinity of massive objects such as BHs. In such conditions, the gravitational field itself acts as a source of electromagnetic fields, amplifying them and, in turn, influencing the curvature of spacetime.

As a consequence, exact solutions for RBHs can be constructed if electric or magnetic charges associated with NLED are non-zero \cite{Bronnikov, FanWang}.

In this regard, various approaches have been discussed in the literature. One prototype involves the utilization of the Born-Infeld Lagrangian, aimed at eliminating the central singularity of a point charge's electromagnetic field as well as its energy divergence \cite{Born}. Following a similar approach, RBH solutions in NLED include those proposed by Bardeen, Hayward, Dymnikova, Fan-Wang, among others. Bardeen introduced the first model of RBHs, now known as the Bardeen BH \cite{Bardeen} \footnote{Subsequently, Ayon-Beato-Garcia  and Bronnikov \cite{Bronnikov} demonstrated that these solutions arise from NLED as a source. They proposed a magnetic monopole as a source within the framework of NLED, leading to the Bardeen BH solution from the Einstein field equations \cite{Beato, 2000, Bronnikov}.}.

Interestingly, Hayward constructed a RBH by assuming a particular mass function \cite{Hayward} where vacuum energy avoids the singularity. Introducing the Lagrangian for NLED to the first order, Bronnikov \cite{Bronnikov}, Berej et al. \cite{Berej} initially constructed the RBH, whereas in Ref. \cite{Dymnikova}, using the concepts put forth by Bronnikov \cite{Bronnikov}, Berej et al. \cite{Berej}, Dymnikova presented an exact, regular, spherically symmetric, charged BH solution\footnote{This solution was constructed using a NLED theory with a Hamiltonian-like function. Balart and Vagenas \cite{Balart} constructed charged RBHs using Einstein-NLED theory. They developed the general lapse function for mass distribution functions based on continuous probability distributions.}. Additionally, Fan and Wang constructed a new class of magnetic solutions with spherical symmetry \cite{FanWang}.

Interestingly, Hayward constructed a RBH by assuming a particular mass function \cite{Hayward}, where vacuum energy prevents the singularity. Introducing the Lagrangian for NLED to the first order, Bronnikov and Berej et al. initially constructed RBHs \cite{Bronnikov, Berej}. In a subsequent work, Dymnikova presented an exact, regular, spherically symmetric, charged BH solution, building upon the concepts proposed by Bronnikov and Berej et al. \cite{Dymnikova}. \footnote{This solution was derived using an NLED theory with a Hamiltonian-like function. Balart and Vagenas also constructed charged RBHs using Einstein-NLED theory, developing the general lapse function for mass distribution functions based on continuous probability distributions \cite{Balart}.}. Additionally, Fan and Wang constructed a new class of magnetic solutions with spherical symmetry \cite{FanWang}.

When NLED is connected to General Relativity and complies with the Weak Energy Condition (WEC), the presence of electrically charged regular structures and a limitation on self-energy that becomes infinite for a point charge are ensured. It is important to note that a field need not be weak to be regular; instead, one should disregard only the requirement of Maxwell's weak field limit at the center, which forms the basis of non-existence theorems. It was demonstrated in \cite{Dymnikova} that there are electrically charged structures with regular centers that exhibit regular geometry, fields, and stress-energy tensors without a Maxwell limit as $r \to 0$.

Moreover, the study of BHs holds significant importance in modern astrophysics. A hot accretion disk, typically formed around a BH by gaseous material \cite{Churilova}, generates a distinct spectrum of electromagnetic radiation. The processes occurring within these accretion disks result in the radiation they produce. Hence, the physics of BHs captivates researchers in testing theories of gravity in regions of strong gravitational fields. Analyzing images of BH accretion provides an opportunity to bridge the gap between theoretical predictions and observational evidence.

This encourages an investigation of the luminosity properties of the accretion disk surrounded by NLED sources for RBHs. Relevant clues have recently been carried forward. For example, in Ref. \cite{Akbarieh}, the authors investigated the non-rotating Bardeen and Hayward BHs. They examined the physical properties of non-rotating RBHs,  differential luminosity, temperature, and efficiency of converting accretion mass into radiation\footnote{Remarkably, the authors showed that the accretion disk around the non-rotating Bardeen and Hayward RBHs emits more energy compared to the Schwarzschild BH. Furthermore, the authors demonstrated that  Bardeen and Hayward RBHs are more efficient than Schwarzschild BHs at converting aggregate mass into radiation.}. Thin accretion disks in the gravitational field of a class of rotating RBHs were examined by the authors in Ref. \cite{2024EPJC...84..230B}. Important accretion disk parameters were determined: disk radiation flux, differential and spectral luminosity, and the radius of the innermost stable circular orbit  ($r_{\text{ISCO}}$). Additionally, the quantification of the mass-to-energy conversion efficiency in accretion disks revealed distinct departures from the Kerr solution predictions.

In this article, following the standard theory of BH accretion developed in Refs. \cite{Novikov,Thorne}, we work with a theoretical model for determining the key features of accretion disks. First, the motion of particles on circular orbits within the Dymnikova and Fan-Wang metrics is analyzed. After that, we determine the metrics' kinematic properties and the innermost stable circular orbits (ISCOs). Additionally, we compute the radiative flux, temperature, and differential luminosity of the accretion disk for both metrics. These computations provide valuable insights into the observable aspects of accretion processes and the radiative emissions from surrounding materials. We compare our results in view of recent developments related to BHs and regular solutions previously investigated. 

The paper is organized as follows. In Sec. \ref{sez2}, we contemplate the class of NLED RBHs. In Sec. \ref{sez3}, we provide a comprehensive review of the thin accretion disk formalism. In Sec. \ref{sez4}, we explore the conversion efficiency of mass to radiation and, finally, in Sec. \ref{sez5} we present conclusions and perspectives\footnote{Throughout the paper, we use natural units, setting $G=c=1$.}. 

\section{Regular solutions from NLED: The Dymnikova and Fan-Wang spacetimes}\label{sez2}

Investigation the structures of RBHs \cite{Zaslavskii, Bonanno, Rubio, Borde}, in analogy to singular BHs, can include dynamics \cite{Miao, Yang, Guo}, thermodynamics \cite{Maluf2018, Wei2018, Huang2023, Sharif2020, Myung2009, Yan-Gang, Hao, Lopez}, shadows \cite{Olmo2023,Tsukamoto2018,Sau2023, Ling2022, Zdenek2019, Ahmed2022, Sushant,Ghosh2020, Kumar, Kraav, Bambi, Abdujabbarov}, the properties of the accretion disk around RBHs \cite{2024EPJC...84..230B,Guo2023, Narzilloev, Martino2023, Zeng, Akbarieh}, as well as quasinormal modes \cite{Nomura2005, Toshmatov, Lemos, Santos2020, Panotopoulos, Lan}, among others. 

Focusing on spherically-symmetric spacetime, we have 
\begin{eqnarray}
\label{eq.01}
ds^2=-f(r)dt^2+\frac{dr^2}{f(r)}+r^2(d\theta^2+\sin^2{\theta}d\phi^2),
\end{eqnarray}
where for NLED, we can first single out the Dymnikova BH \cite{Dymnikova, 2022GReGr..54..148W}:
\begin{eqnarray}
\label{eq.03}
   f_{D}(r)=1-\frac{4m}{\pi r}\left(\arctan\Big(\frac{r}{l_{D}}\Big)-\frac{rl_{D}}{r^2+l_{D}^2}\right),
\end{eqnarray}\normalsize 
where $l_D$ is the length scale, $l_{D}=\frac{\pi q_{D}^2}{8m}$, $m$ is the total mass and $q_D$ is the electric charge for the Dymnikova BH.

Additionally, another possible solution within the realm of NLED is offered by the Fan-Wang spacetime, having \cite{FanWang}:
{\begin{eqnarray}
\label{eq.04}
f_{FW}(r)=1-\frac{2mr^2}{(r+l_{FW})^3},
\end{eqnarray}}\normalsize
where $l_{FW}$ is the magnetic charge parameter for the Fan-Wang BH. Hereafter, the indices ``D"  and ``FW" will stand for Dymnikova and  Fan-Wang, respectively.

The Schwarzschild solution is recovered in the limit of either $l_{D}, l_{FW}\rightarrow 0$, whereas for the Dymnikova and Fan-Wang BHs the event horizons imply the conditions 
\begin{eqnarray}
\label{eq.05}
  \frac{\pi r}{4m}-\arctan\Big(\frac{r}{m \bar{l}_{D}}\Big)-\frac{m r \bar{l}_{D}}{r^2+ m^2 \bar{l}_{D}^2}=0,
\end{eqnarray}\normalsize 
and
\begin{eqnarray}
\label{eq.06}
  (r+m\bar{l}_{FW})^{3}-2m r^2=0,
\end{eqnarray}\normalsize
where $\bar{l}_{D}=l_{D}/m$ and $\bar{l}_{FW}=l_{FW}/m$ are the normalized charges. For both  Dymnikova and Fan-Wang BHs, the event horizons exist only when $\bar{l}_{D}\leq0.452$ and $\bar{l}_{FW}\leq8/27$, respectively. 
\begin{figure*}
    \centering
   {\includegraphics[width=0.45\linewidth]{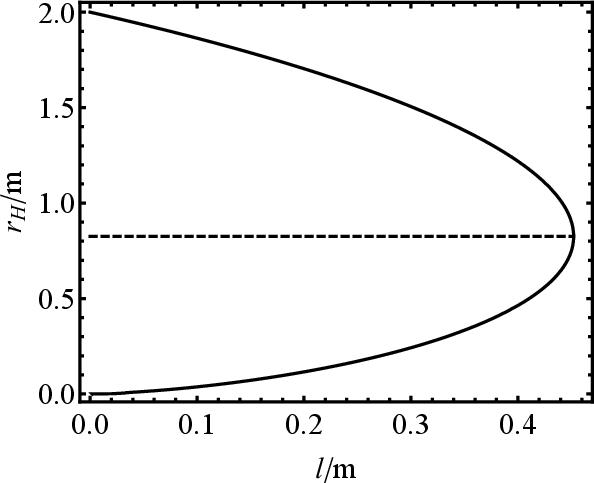} }%
    \qquad
    {\includegraphics[width=0.45\linewidth]{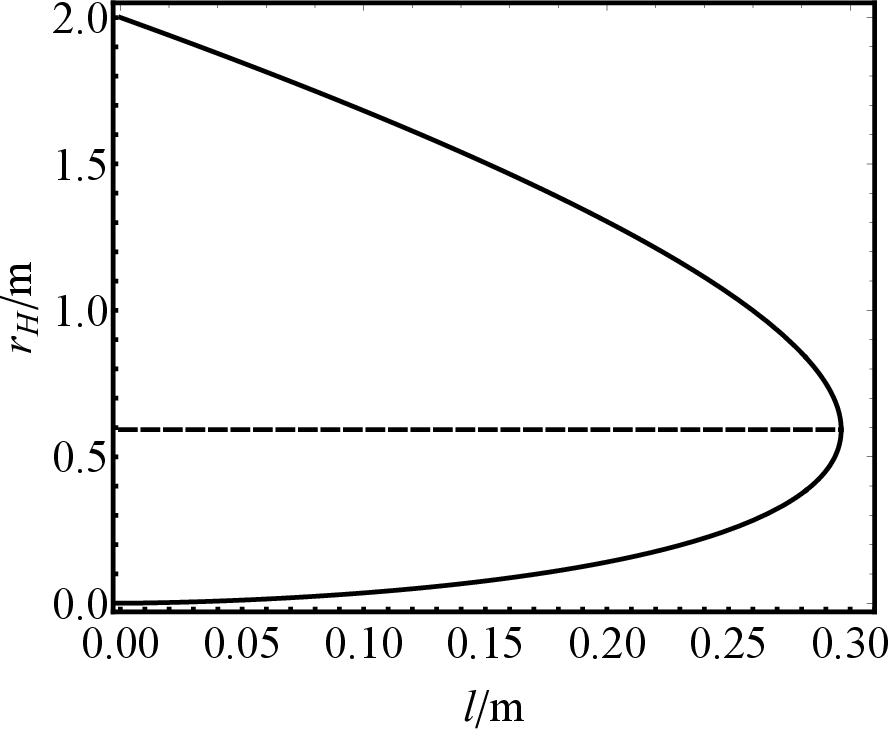} }%
    \caption{The inner and outer horizons radii normalized by the mass $r_H/m$ versus $l/m$. The dashed black line indicates the boundary between inner and outer horizons. on the left, we have the Dymnikova BH, where at the border $r_H/m=0.82532$ and $l/m=0.452167$. On the right, we have the Fan-Wang BH, the boundary is at $r_H/m=16/27\approx0.59259$ and $l/m=8/27\approx0.29629$.}%
    \label{fig:rh_FW_Dym}%
\end{figure*}

\subsection{Numerical analysis of the solutions}

In Fig.~\ref{fig:rh_FW_Dym}, we present the radii of inner and outer horizons normalized by mass versus the parameters of the model also normalized by mass. The left panel illustrates the horizon radii for the Dymnikova BH, while the right panel depicts the horizon radii for the Fan-Wang BH. The dashed black line in both panels marks the boundary between the inner and outer horizons. At this boundary, the values of $r_H/m$ and $l/m$ correspond to the extreme BH cases. As $l/m$ approaches zero, the inner horizon $r_H/m$ tends to 0, while the outer horizon tends to 2 in both models, as expected.

\begin{figure*}[ht]
\begin{minipage}{0.49\linewidth}
\center{\includegraphics[width=0.98\linewidth]{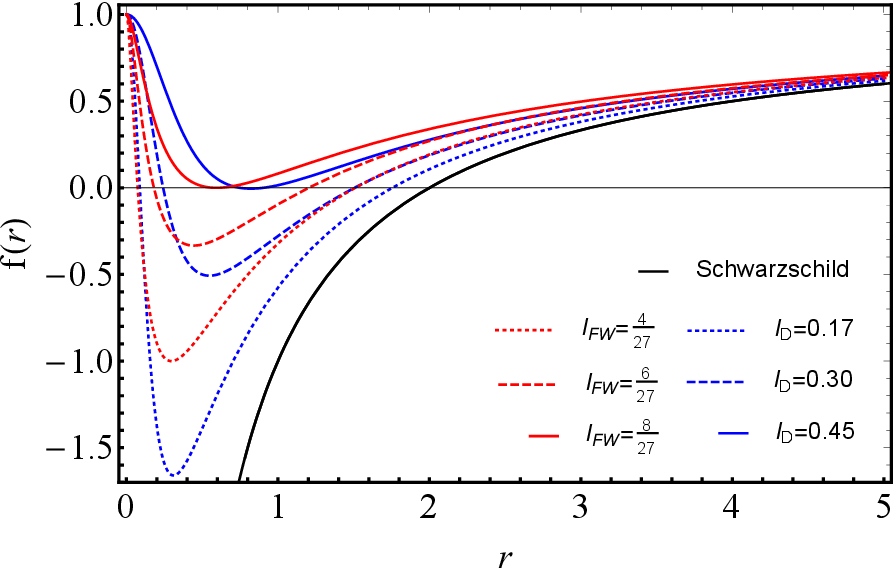}\\ }
\end{minipage}
\hfill 
\begin{minipage}{0.50\linewidth}
\center{\includegraphics[width=0.97\linewidth]{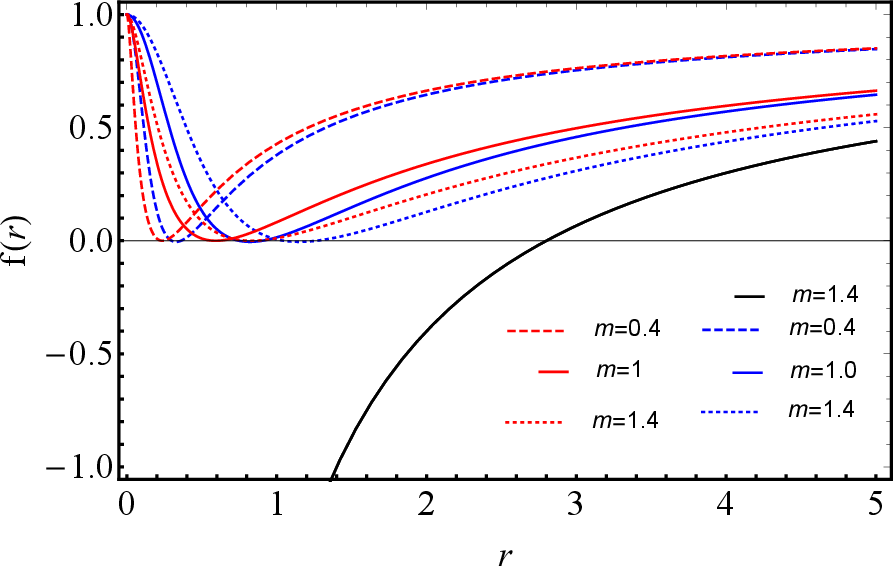}\\ }
\end{minipage}
\caption{Left panel: The metric coefficient $f(r)$ for the Dymnikova and Fan-Wang BHs with different values of the free parameter $l_{D}$ and $l_{FW}$. Right panel: The metric coefficient $f(r)$ plot for Dymnikova and Fan-Wang BHs with varying mass $m$ and constant free parameter values $l_{D}=0.452$ and $l_{FW}=8/27$.}
\label{fig:fr}
\end{figure*}

In Fig. \ref{fig:fr} (left panel), we plotted $f(r)$ against $r$ for the Dymnikova and Fan-Wang BHs with varying free parameter values. The Dymnikova BH is depicted in blue, and the Fan-Wang BH is represented in red. The blue and red solid curves indicate critical values of the free parameters, while the blue and red dashed and dotted curves correspond to values below the critical threshold. For comparison, the function $f(r)$ for the Schwarzschild metric with $m=1$ is shown as the black solid curve. Fig. \ref{fig:fr} (left panel) shows that, in comparison to the Schwarzschild BH, the event horizons of RBHs are shifted to lower values. Both the  Dymnikova and Fan-Wang BHs exhibit one event horizon at critical values ${l}_{D}=0.452$ and ${l}_{FW}=8/27$. The blue (Dymnikova solution) and red (Fan-Wang solution) dashed curves represent $f(r)$ for the free parameters ${l}_{D}=0.30$ and ${l}_{FW}=6/27$, which are below the critical values. Additionally, the blue and red dotted curves correspond to the values of ${l}_{D}=0.17$ and ${l}_{FW}=4/27$, which are also below the critical values. Furthermore, similar to the Hayward and Bardeen BHs \cite{Akbarieh}, the function $f(r)$ tends towards $1$ as $r\rightarrow 0$. Additionally, for $r\rightarrow\infty$, the behavior of Dymnikova and Fan-Wang BHs approaches that of a Schwarzschild BH (see Fig. \ref{fig:fr} left panel).

In Fig.\ref{fig:fr} (right panel), we plot $f(r)$ as a function of $r$ for Dymnikova and Fan-Wang BHs with varying mass $m$ and constant values
of the free parameter $\bar {l}_{D}=0.45$, $\bar {l}_{FW}=8/27$.

\subsection{Morphological structures of the solutions}

In order to make our computation more compact, we can take into account an effective mass representation, $M(r)$ as a function of the radial coordinate, rewriting Eqs. (\ref{eq.03}) and (\ref{eq.04}) as 

{\begin{eqnarray}
\label{eq.07}
ds^2 = -\left(1-\frac{2M(r)}{r}\right)dt^2+\frac{dr^{2}}{\left(1-\frac{2M(r)}{r}\right)}+r^2d\Omega^2,
\end{eqnarray}}\normalsize
as inspired by Ref. \cite{quasidamour}, following the Damour-Solodukhin description and, so, having for the Dymnikova and Fan-Wang, respectively 
\begin{eqnarray}
       M_{D}(r)&=&\frac{2m}{\pi}\left(\arctan\Big(\frac{r}{l_D}\Big)-\frac{r {l}_{D}}{r^2+ l_{D}^2}\right),\\ 
       M_{FW}(r)&=&\frac{m r^3}{(r+l_{FW})^{3}}.
\end{eqnarray}    

Lying on the equatorial plane, $\theta=\pi/2$, the Lagrangian reads 
$2\mathcal{L}=-f(r)\dot{t}^2+\frac{\dot{r}^2}{f(r)}+r^2\dot{\phi}^2$, where dot represents the derivative with respect to the affine parameter. Accordingly, the generalized momenta are
\begin{eqnarray}
\label{eq.11}
   p_{t}&=&\frac{\partial\mathcal{L}}{\partial\dot{t}}=-\Big(1-2\frac{M(r)}{r}\Big)\dot{t}=-E,\\
\label{eq.12}
   p_{r}&=&\frac{\partial\mathcal{L}}{\partial\dot{r}}=\Big(1-2\frac{M(r)}{r}\Big)^{-1}\dot{r},\\
\label{eq.13}
   p_{\phi}&=&\frac{\partial\mathcal{L}}{\partial\dot{\phi}}=r^2\dot{\phi}=L.
\end{eqnarray}
where, $E$ and $L$ are the energy and angular momentum per unit of the particle's rest mass, respectively. The Hamiltonian is 
\begin{eqnarray}
\label{eq.14}
2H=-E\dot{t}+\Big(1-2\frac{M(r)}{r}\Big)^{-1}\dot{r}^2+L\dot{\phi}=-1.
\end{eqnarray}
The energy equation can be obtained by solving Eqs.~(\ref{eq.11}) and (\ref{eq.13}) for $\dot{t}$, $\dot{\phi}$ and then combining the results within Eq. (\ref{eq.14}),
\begin{eqnarray}
\label{eq.15}
\frac{1}{2}\dot{r}^2+V_{eff}(r)=\frac{1}{2}\Big(E^2-1\Big),
\end{eqnarray}
where the effective potential per unit mass acquires the form,
\begin{eqnarray}
\label{eq.16}
V_{eff}(r)=-\frac{M(r)}{r}+\frac{L^2}{2r^2}-\frac{M(r)L^2}{r^3}.
\end{eqnarray}

It is helpful to write $r=r(\phi)$ \cite{Yakov}, thus having from Eq.~\ref{eq.13},
\begin{eqnarray}
\label{eq.17}
\dot{r}=\frac{dr}{d\tau}=\frac{dr}{d\phi}\frac{d\phi}{d\tau}=\frac{L}{r^2}\frac{dr}{d\phi},
\end{eqnarray}
and so, substituting Eq. (\ref{eq.17}) in Eq. (\ref{eq.15}), we obtain

\begin{eqnarray}
\label{eq.18}
\Big(\frac{du}{d\phi}\Big)^2+u^2=\frac{E^2-1}{L^2}+\frac{2uM(u)}{L^2}+2u^3M(u),
\end{eqnarray}
where  $r=1/u$. 

Further, replacing the mass profiles for the two solutions, and then taking the derivative with respect to $\phi$ of Eq. (\ref{eq.18}), we infer 

\begin{widetext}
\begin{eqnarray}
\label{eq.19}
\frac{d^2u}{d\phi^2}+u &=& \frac{2m}{\pi\Big(1+u^2l_D^2\Big)^2}\Bigg[\frac{1}{L_D^2}\Bigg(\Big(1+u^2l_D^2\Big)^2 \arctan\Big(\frac{1}{u l_D}\Big)-ul_D\Big(3+u^2l_D^2\Big)\Bigg)+\nonumber\\
&+&3u^2\Big(1+u^2l_D^2\Big)^2\arctan \Big(\frac{1}{ul_D}\Big)-u^3 l_D \Big(5+3u^2 l_D^2\Big)\Bigg].
\end{eqnarray}\normalsize 

 \end{widetext}
 and 
 \begin{eqnarray}
\label{eq.20}
\frac{d^2u}{d\phi^2}+u=\frac{m}{\Big(1+ul_{FW}\Big)^{4}}\Bigg[\frac{1}{L_{FW}^2}\Big(1-2ul_{FW}\Big)+3u^2\Bigg],
\end{eqnarray}\normalsize 
for the Dymnikova and Fan-Wang spacetimes, respectively. 

For circular orbits in the equatorial plane, $\dot{r}=0$, i.e., showing  that $r$ and $u = 1/r$ are constant. As a byproduct, we  calculate the angular momenta using Eqs.~\eqref{eq.19}  and \eqref{eq.20}, having

\begin{widetext}
\begin{eqnarray}
\label{eq.21}
L_D&=&\sqrt{2}mx\sqrt{\frac{\Big(x^2+\bar{l}_D^2\Big)^2\arctan\left(\frac{x}{\bar{l}_D}\right)-x \bar{l}_D \Big(3x^2+\bar{l}_D^2\Big)}{x\Big(6\bar{l}_D^3+\pi \bar{l}_D^4+10x^2 \bar{l}_D+2\pi x^2 \bar{l}_D^2+\pi x^4\Big)-6\Big(x^2+\bar{l}_D^2\Big)^2\arctan\left(\frac{x}{\bar{l}_D}\right)}},
\end{eqnarray}\normalsize
 \end{widetext}
and 
\begin{eqnarray}
\label{eq.22}
L_{FW}&=&mx^2\sqrt{\frac{x-2\bar{l}_{FW}}{\Big(x+\bar{l}_{FW}\Big)^{4}-3x^3}},
\end{eqnarray}\normalsize 
again for the Dymnikova and Fan-Wang spacetimes, being conveniently calculated in dimensionless values $u=1/mx$. Additionally, the specific energies for the two cases read

\begin{widetext}
\begin{eqnarray}
\label{eq.23}
E_D&=&\frac{x\Big(4\bar{l}_D+\pi \bar{l}_D^2+\pi x^2\Big)-4\Big(x^2+\bar{l}_D^2\Big)\arctan\left(\frac{x}{\bar{l}_D}\right)}{\sqrt{{\pi x^2\Big(6\bar{l}_D^3+\pi \bar{l}_D^4+10x^2 \bar{l}_D+2\pi x^2 \bar{l}_D^2+\pi x^4\Big)-6\pi x\Big(x^2+\bar{l}_D^2\Big)^2\arctan\left(\frac{x}{\bar{l}_D}\right)}}},
\end{eqnarray}\normalsize
 \end{widetext}
and
\begin{eqnarray}
\label{eq.24}
E_{FW}=\frac{\Big(x+\bar{l}_{FW}\Big)^{3}-2x^2}{\Big(x+\bar{l}_{FW}\Big)\sqrt{\Big(x+\bar{l}_{FW}\Big)^{4}-3x^3}},
\end{eqnarray}\normalsize

The angular velocity  for the Dymnikova and Fan-Wang BHs acquires the form

\begin{widetext}
\begin{eqnarray}
\label{eq.25}
\Omega_D=\frac{\dot{\phi}}{\dot{t}}=\frac{1}{mx}\sqrt{\frac{2}{\pi x}\arctan\left(\frac{x}{\bar{l}_D}\right)- \frac{2\bar{l}_D \Big(3x^2+\bar{l}_D^2\Big)}{\pi\Big(x^2+\bar{l}_D^2\Big)^2}}, 
\end{eqnarray}\normalsize
\end{widetext}
 and 
\begin{eqnarray}
\label{eq.26}
\Omega_{FW}&=&\frac{\dot{\phi}}{\dot{t}}=\frac{\sqrt{x-2\bar{l}_{FW}}}{m\Big(x+\bar{l}_{FW}\Big)^{2}}.
\end{eqnarray}\normalsize

The effective potentials for Dymnikova and Fan-Wang BHs can be found by substituting Eqs.  (\ref{eq.21}) and (\ref{eq.22}) into Eq. (\ref{eq.16}), respectively.
\begin{widetext}
\begin{eqnarray}
\label{eq.27}
V_{eff-D}=\frac{x^2\bar{l}_D\Big(8\bar{l}_D+\pi \bar{l}_D^2-\pi x^2\Big)-x\Big(x^2+\bar{l}_D^2\Big)\Big(16\bar{l}_D+\pi \bar{l}_D^2+\pi x^2\Big)\arctan\left(\frac{x}{\bar{l}_D}\right)+8\Big(x^2+\bar{l}_D^2\Big)^2\arctan\left(\frac{x}{\bar{l}_D}\right)^2}{{\pi x^2\Big(6\bar{l}_D^3+\pi \bar{l}_D^4+10x^2 \bar{l}_D+2\pi x^2 \bar{l}_D^2+\pi x^4\Big)-6\pi x\Big(x^2+\bar{l}_D^2\Big)^2\arctan\left(\frac{x}{\bar{l}_D}\right)}},
\end{eqnarray}\normalsize
\end{widetext}
and
\begin{eqnarray}
\label{eq.28}
V_{eff-FW}&=&\frac{x^2\Big[4x^2-\Big(x+4\bar{l}_{FW}\Big)\Big(x+\bar{l}_{FW}\Big)^2\Big]}{2\Big(x+\bar{l}_{FW}\Big)^2\Big[\Big(x+\bar{l}_{FW}\Big)^4-3x^3\Big]}.
\end{eqnarray}

At this stage, the dimensionless ISCO radii, $x_{ISCO}$, may be computed from \cite{Thorne}, through the definition,  
$\frac{d^2V_{eff}}{dx^2}=0$.

Thus, the derivative of the effective potential $V_{eff}$ is computed \cite{Harko,Kovacs}
\begin{widetext}
\begin{eqnarray}
    \label{eq.30}
\frac{d^2V_{eff}}{dx^2}&=& \frac{4x\bar{l}_{D}\Big(\Big(x^2+\bar{l}_D^2\Big)^2+4x^4\Big)-4\Big(x^2+\bar{l}_D^2\Big)^3\arctan\left(\frac{x}{\bar{l}_D}\right)}{\pi x^3 \Big(x^2+\bar{l}_D^2\Big)^3}+\nonumber\\
&+&\bar{L^2}_{D} \Bigg(\frac{3}{x^4}+ \frac{x \bar{l}_D \Big(\Big(x^2+\bar{l}_D^2\Big)^2 + 16 \Big(2x^2+\bar{l}_D^2\Big)\Big)-24\Big(x^2+\bar{l}_D^2\Big)^3\arctan\left(\frac{x}{\bar{l}_D}\right)}{\pi x^5\Big(x^2+\bar{l}_D^2\Big)^3}\Bigg),
\end{eqnarray}
\end{widetext}
and 

\begin{equation}
\label{eq.31}
\begin{split}
\frac{d^2V_{eff}}{dx^2}=&\frac{12x\bar{l}_{FW}-2(x+\bar{l}_{FW})^2}{(x+\bar{l}_{FW})^5}\\
&+3\bar{L^2}_{FW}\Big(\frac{1}{x^4}-\frac{4}{(x+\bar{l}_{FW})^5}\Big),
\end{split}
\end{equation}

where for brevity $\bar{L}\equiv L/m$. So, by inserting Eqs. (\ref{eq.21}) and (\ref{eq.22}) in Eqs.(\ref{eq.30}) and (\ref{eq.31}), we get

\begin{widetext}
\begin{eqnarray}
 \label{eq.32}
\frac{d^2V_{eff}}{dx^2} & =& \frac{2x\Big(24\bar{l}_D\Big(2x^4+\bar{l}_D^4\Big)+104x^2\bar{l}_D^3+ \pi\Big(x^2+\bar{l}_D^2\Big)^3\Big) \arctan\left(\frac{x}{\bar{l}_D}\right)-2x^2\bar{l}_D \Big(12\bar{l}_D^3 +\pi \bar{l}_D^4+68x^2\bar{l}_D+8\pi x^2\bar{l}_D^2-\pi x^4 \Big)}{\pi x^3 \Big(x^2+\bar{l}_D^2\Big)\Big(x\Big(6\bar{l}_D^3+\pi \bar{l}_D^4+10x^2 \bar{l}_D+2\pi x^2 \bar{l}_D^2+\pi x^4\Big)-6\Big(x^2+\bar{l}_D^2\Big)^2\arctan\left(\frac{x}{\bar{l}_D}\right)\Big)}-\nonumber\\
&-&\frac{24\Big(x^2+\bar{l}_D^2\Big)^2\arctan\left(\frac{x}{\bar{l}_D}\right)^2}{\pi x^3\Big(x\Big(6\bar{l}_D^3+\pi \bar{l}_D^4+10x^2 \bar{l}_D+2\pi x^2 \bar{l}_D^2+\pi x^4\Big)-6\Big(x^2+\bar{l}_D^2\Big)^2\arctan\left(\frac{x}{\bar{l}_D}\right)\Big)},
\end{eqnarray}

\end{widetext}
and
\begin{equation}
\begin{split}
\label{eq.33}
\frac{d^2V_{eff}}{dx^2}=&\frac{12x\bar{l}_{FW}-2\Big(x+\bar{l}_{FW}\Big)^2}{\Big(x+\bar{l}_{FW}\Big)^5}\\
&+3\frac{\Big(x-2\bar{l}_{FW}\Big)\Big[\Big(x+\bar{l}_{FW}\Big)^5-4x^4\Big]}{\Big(x+\bar{l}_{FW}\Big)^5\Big[\Big(x+\bar{l}_{FW}\Big)^4-3x^3\Big]},
\end{split}
\end{equation}
The $x_{ISCO}$ value for the free parameters' critical values $\bar{l}_{D}=0.45$, $\bar{l}_{FW}=\frac{8}{27}$ one can to compute for the Dymnikova and Fan-Wang BHs using Eqs. (\ref{eq.32}) and (\ref{eq.33}).
 
\begin{eqnarray}
\label{eq.024}
\frac{dL}{dx}=\frac{dE}{dx}=0.
\end{eqnarray}

\begin{figure*}[ht]
\begin{minipage}{0.49\linewidth}
\center{\includegraphics[width=0.985\linewidth]{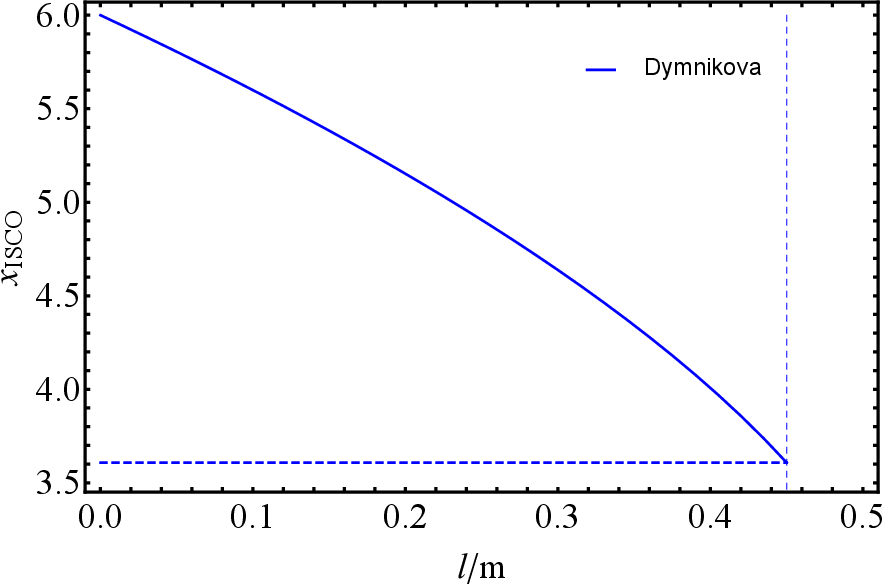}\\ }
\end{minipage}
\hfill 
\begin{minipage}{0.50\linewidth}
\center{\includegraphics[width=0.97\linewidth]{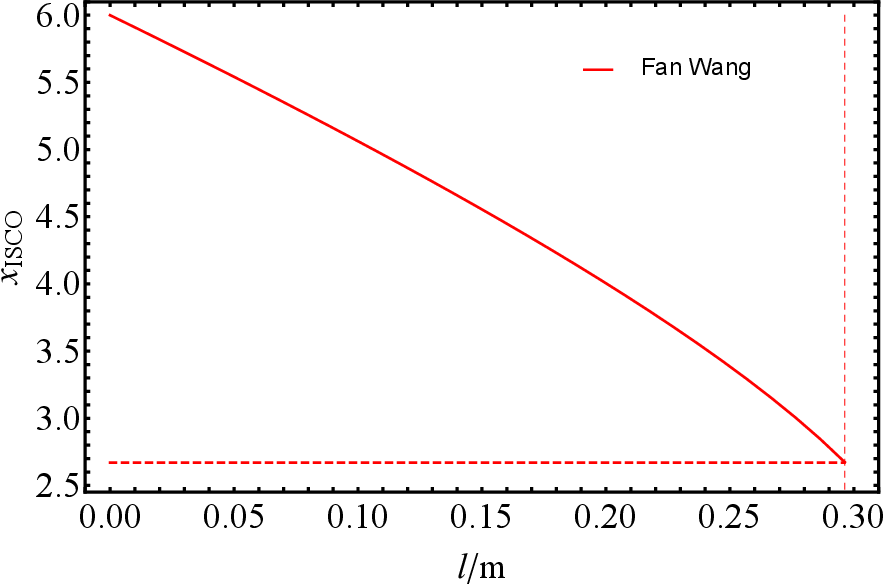}\\ }
\end{minipage}
\caption{Dependence of the $x_{ISCO}$ on the free parameter $l/m$ . Left panel: for the Dymnikova RBH. Right panel: for the Fan-Wang RBH.}
\label{fig:xisco}
\end{figure*}

The $x_{ISCO}$ dependence on the free parameter $\bar{l}$ for the Dymnikova BH (left panel) and Fan-Wang BH (right panel) is shown in Fig. \ref{fig:xisco}.

\begin{figure*}[ht]
\begin{minipage}{0.49\linewidth}
\center{\includegraphics[width=0.98\linewidth]{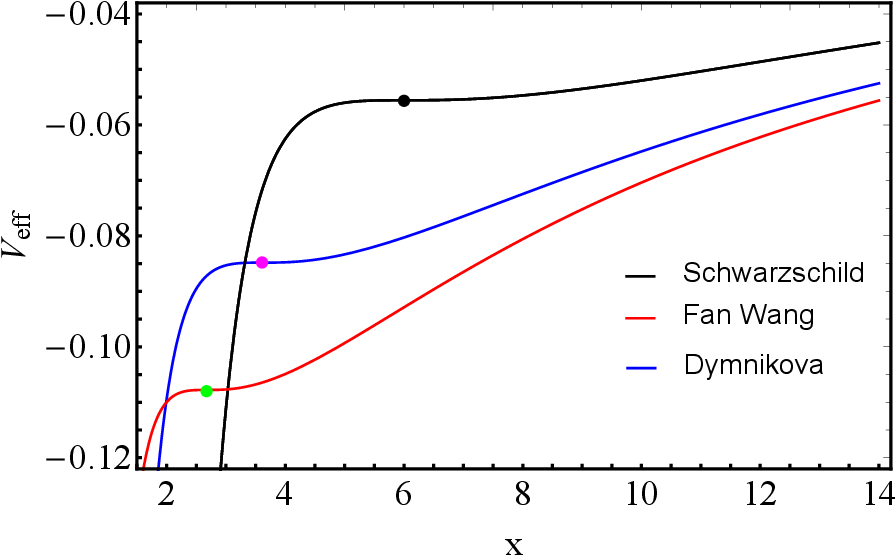}\\ }
\end{minipage}
\hfill 
\begin{minipage}{0.50\linewidth}
\center{\includegraphics[width=0.97\linewidth]{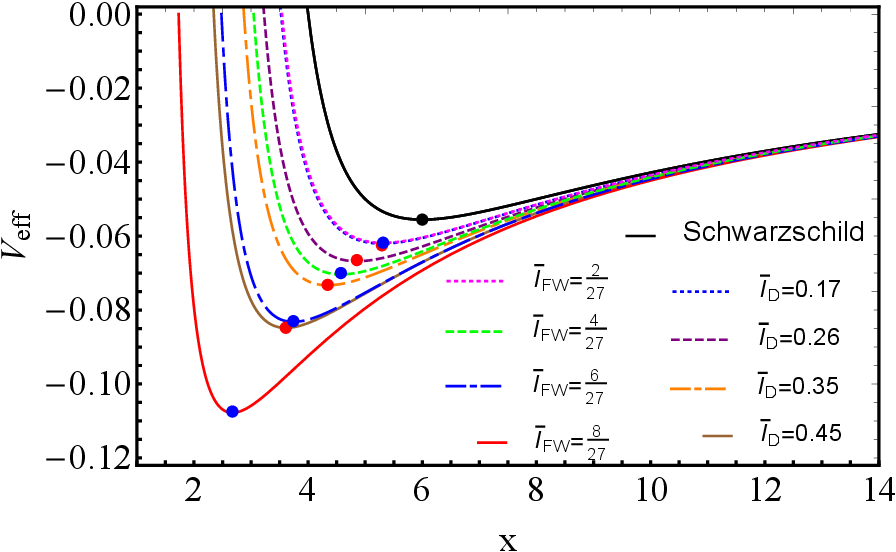}\\ }
\end{minipage}
\caption{Left panel: The Effective potential for the values of $\bar{L}=L/m$ when evaluated at the ISCO versus of $x$ for the Dymnikova BH with $\bar{l}_{D}=0.45$ and the Fan-Wang BH with $\bar{l}_{FW}=8/27$.  Right panel: The effective potential as a function of $x$ for the Dymnikova and Fan-Wang BHs.}
\label{fig:Veff}
\end{figure*}

The effective potential for the constant value of $\bar{L}=L/m$ as determined in the ISCO for the free parameters of the Dymnikova (blue) and Fan-Wang (red) is shown in Figure \ref{fig:Veff} (left panel). The locations of the ISCO for the Schwarzschild, Dymnikova, and Fan-Wang BHs are shown by the black, magenta, and green dots, respectively. 

Additionally, we compared the Dymnikova and Fan-Wang BHs with the Schwarzschild BH by plotting the effective potential  $V_{eff}$  as a function of $x$ in Fig.\ref{fig:Veff} (right panel). As seen from Fig.~\ref{fig:Veff} for both panels the $x_{ISCO}$ and the minimum values of the effective potential move to smaller values. The values of $x_{ISCO}$ for the Schwarzschild, Dymnikova, and Fan-Wang BHs are shown by the black, red, and blue dots, respectively.\\

\begin{figure*}[ht]
\begin{minipage}{0.49\linewidth}
\center{\includegraphics[width=0.98\linewidth]{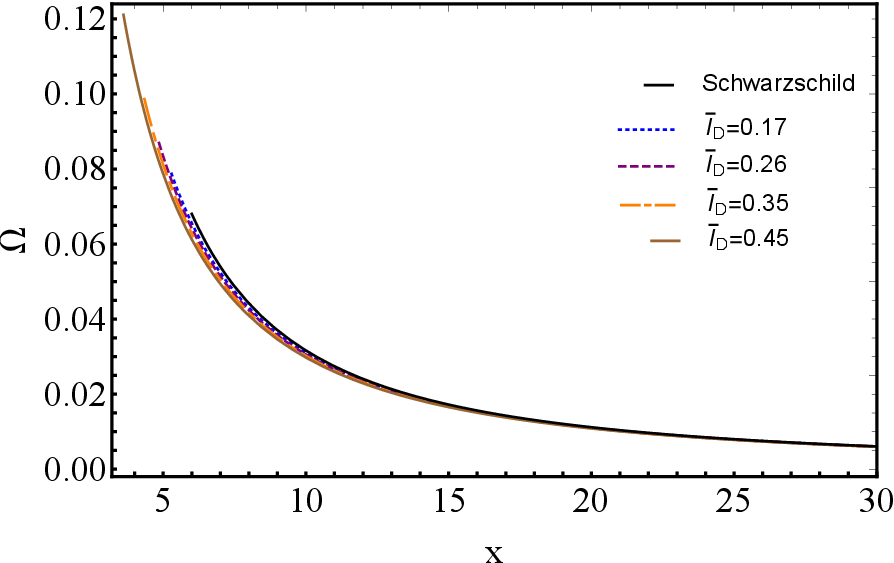}\\ }
\end{minipage}
\hfill 
\begin{minipage}{0.50\linewidth}
\center{\includegraphics[width=0.97\linewidth]{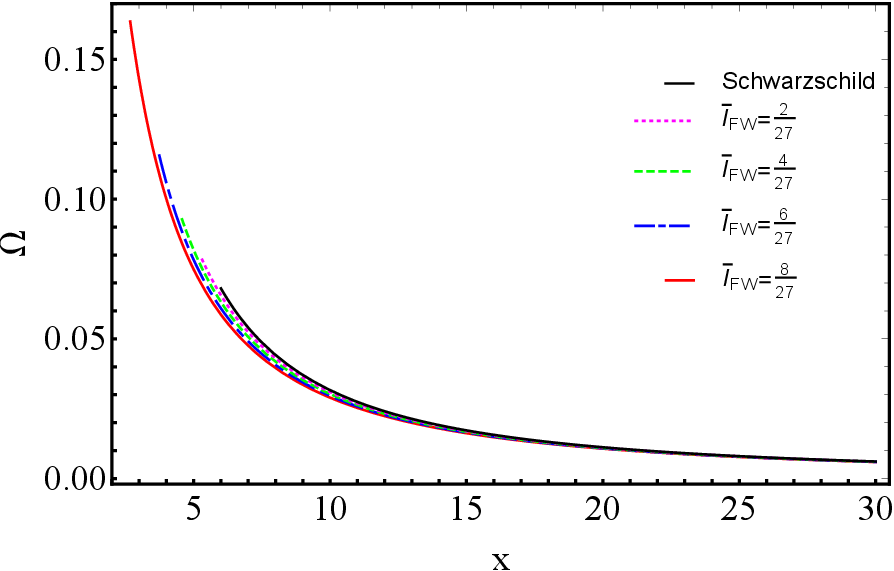}\\ }
\end{minipage}
\caption{Left panel: Angular velocity of test particles versus $x$ for Dymnikova space-time. Right panel: Angular velocity of test particles versus $x$ for Fan-Wang space-time.}
\label{fig:OmegaFWD}
\end{figure*}
In Fig.~\ref{fig:OmegaFWD} we show the orbital angular velocity $\Omega$ of test particles as a function of $x$ Dymnikova (left panel) and Fan-Wang (right panel) spacetimes. In both panels, we see that for different values of $\bar{l}_{D},\bar{l}_{FW}$, the angular velocity is always less than the Schwarzschild BH (black solid curve). 

\begin{figure*}[ht]
\begin{minipage}{0.49\linewidth}
\center{\includegraphics[width=0.99\linewidth]{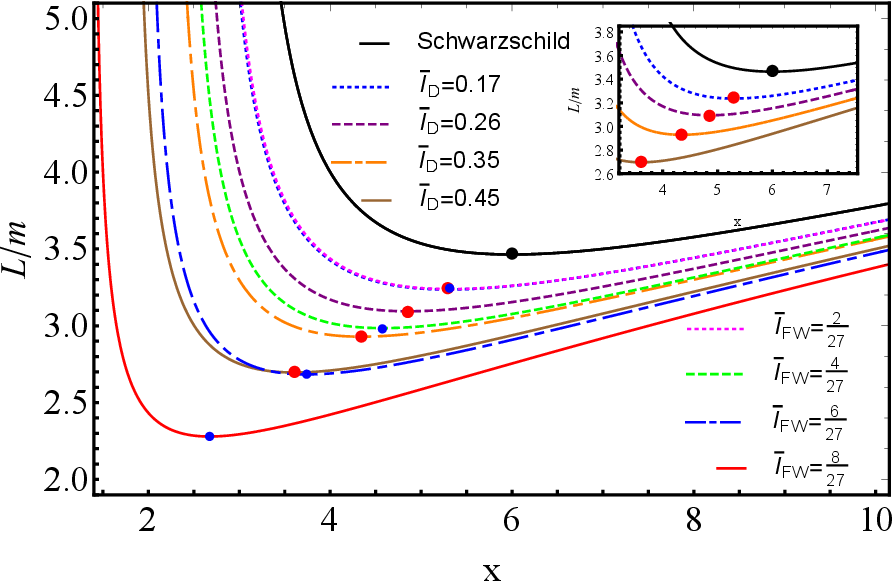}\\ }
\end{minipage}
\hfill 
\begin{minipage}{0.50\linewidth}
\center{\includegraphics[width=0.95\linewidth]{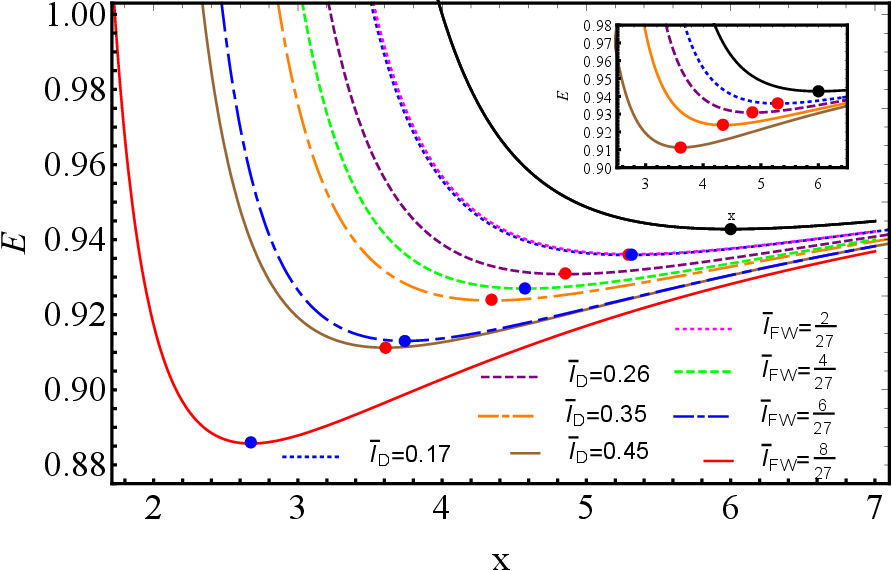}\\ }
\end{minipage}
\caption{Left panel:Angular momentum $\bar{L}=L/m$ as a function of $x$. Right panel: Specific energy $E$ as a function of $x$.}
\label{fig:LE}
\end{figure*}

In Fig. \ref{fig:LE} (left panel), the angular momentum $\bar{L}$ is depicted  as a function of the $x$ for Dymnikova and Fan-Wang spacetimes and compared with the Schwarzschild BH. It is evident from Fig. \ref{fig:LE} that the angular momentum has also been marginally moved to lower values in the $x_{isco}$ region.

The same pattern as the angular momentum is replicated in the curve of specific energy $E$ in terms of $x$, as seen in Fig. \ref{fig:LE},(right panel). The specific energy $E$ of RBHs is somewhat skewed towards lesser values when compared with the Schwarzschild BH, as seen in Fig. \ref{fig:LE},(right panel), which is presented for the critical values of the free parameters $\bar{l}_{D}$ and $\bar{l}_{FW}$ (See Tab. \ref{tab:1}). 

\begin{figure}
\centering
\includegraphics[width=0.97\linewidth]{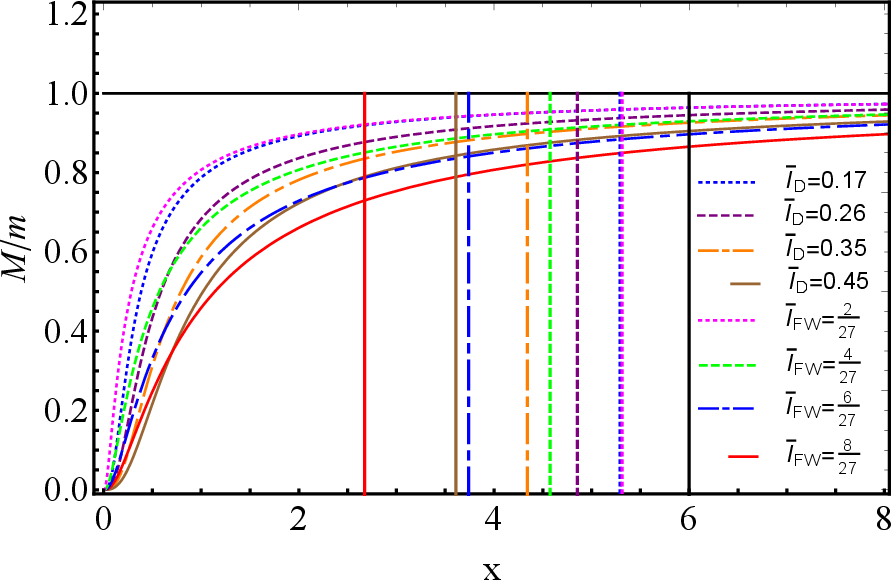}
\caption{The mass curves as a function of $x$ for the Dymnikova and Fan-Wang BHs.} 
\label{fig:Meff}
\end{figure}

In Fig.~\ref{fig:Meff} we compare the mass curve of the Dymnikova and the Fang-Wang BHs as a function of $x$ with respect to the Schwarzschild BH (horizontal black line). The vertical lines show the ISCO for the Dymnikova and Fan-Wang BHs with the free parameters $\bar{l}_{D}$ and $\bar{l}_{FW}$. The horizontal solid line corresponds to the Schwarzschild space-time. As can be seen from Fig.~\ref{fig:Meff}, for a given value of the free parameters $\bar{l}_{D}$ and $\bar{l}_{FW}$, the particle falling into the BH experiences less gravitational pull due to the decreasing mass profile as the $x$ decreases (See. Fig. \ref{fig:Meff}).

\section{Relativistic Characteristics of Thin Accretion Disks}\label{sez3}

The derivative of the luminosity at infinity $\mathcal{L}_{\infty}$ is obtained by combining the laws of energy conservation and angular momentum \cite{Thorne,Joshi}

\begin{eqnarray}
\label{eq.025}
\frac{d\mathcal{L}_{\infty}}{dlnr}=4\pi r\sqrt{-g}E\mathcal{F}(r),
\end{eqnarray}

The radiant energy flux $\mathcal{F}$ emitted from the disk's surface is determined by
\begin{eqnarray}
   \label{eq.026}
\mathcal{F}(r)=-\frac{\dot{{\rm m}}}{4\pi \sqrt{-g}} \frac{\Omega_{,r}}{\left(E-\Omega L\right)^2 }\int^r_{r_{i}} \left(E-\Omega L\right) L_{,\tilde{r}}d\tilde{r}, 
\end{eqnarray}
where $\dot{m}$ is the mass accretion rate of the disk, which is assumed to be constant, $r_i=r_{ISCO}$ and $\sqrt{-g}=r$ applies to the Dymnikova, Fan-Wang, and  Schwarzschild spacetimes. 

If the accretion disk is assumed to be in thermodynamic equilibrium, then its surface radiation can be considered blackbody radiation with a temperature determined by \cite{Zuluaga}

\begin{eqnarray}
\label{eq.027}
{T}(r)=\sigma^{-\frac{1}{4}}\mathcal{F}(r)^{\frac{1}{4}},
\end{eqnarray}
where $\sigma$ is the Stefan-Boltzmann constant.

\begin{figure*}[ht]
\begin{minipage}{0.49\linewidth}
\center{\includegraphics[width=0.97\linewidth]{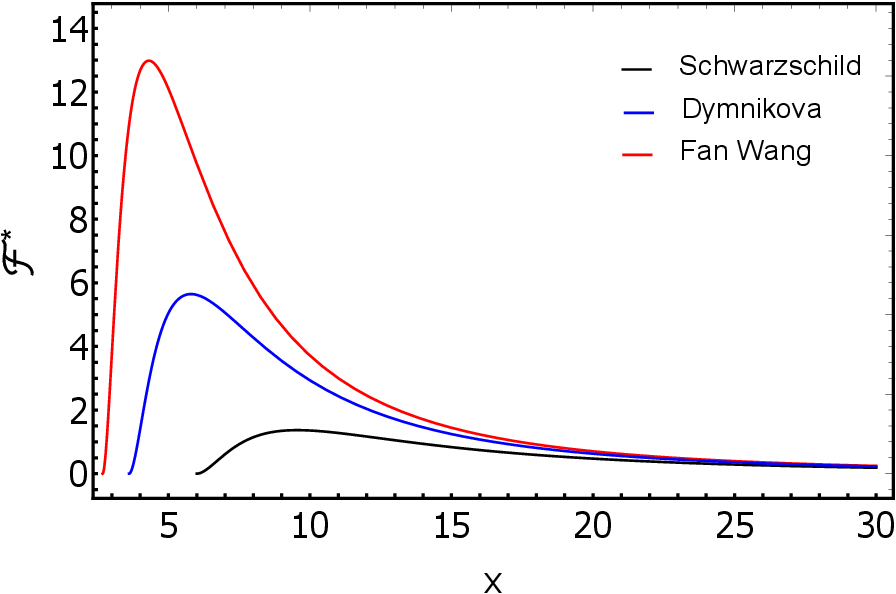}\\ }
\end{minipage}
\hfill 
\begin{minipage}{0.50\linewidth}
\center{\includegraphics[width=0.97\linewidth]{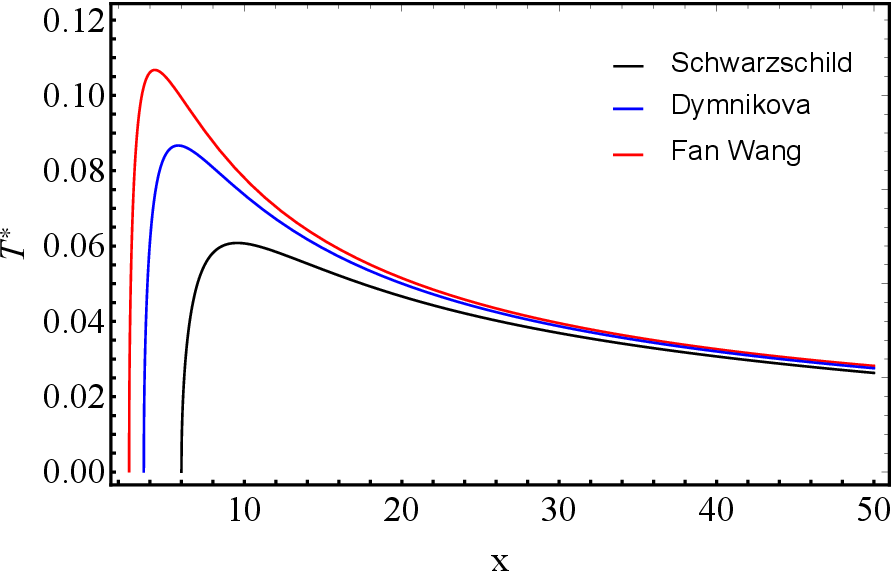}\\ }
\end{minipage}
\caption{Left panel: Radiative flux $\mathcal{F}^*$ of the accretion disk multiplied by $10^5$ versus  $x$.  Right panel: The temperature $T^{*}$ in the accretion rate unit of the disk around the RBHs.}
\label{fig:FluxT}
\end{figure*}

\begin{figure}
\centering
\includegraphics[width=0.97\linewidth]{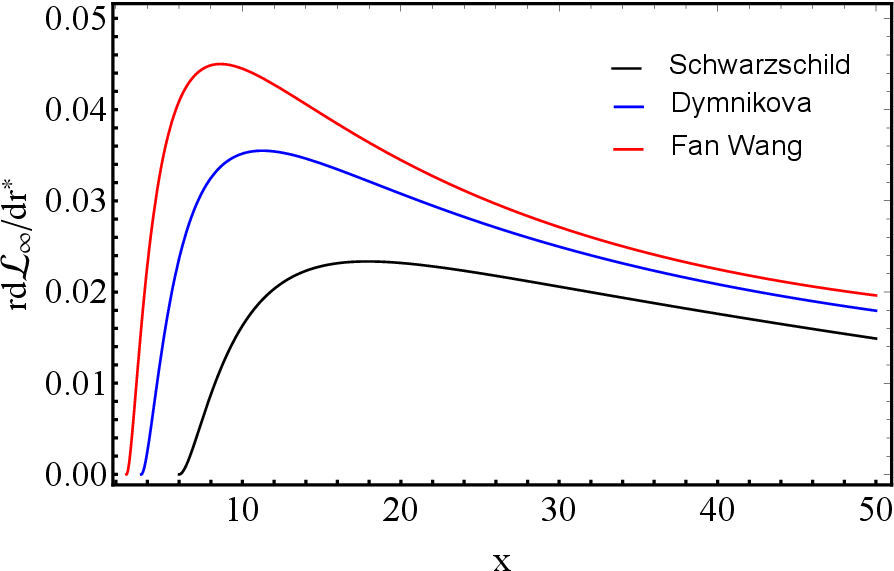}
\caption{Differential luminosity of the thin accretion disk around RBHs.} 
\label{fig:Diflum}
\end{figure}

In Figs.~\ref{fig:FluxT} and \ref{fig:Diflum} we plotted the flux, the temperature of the accretion disk in the accretion rate unit, and the differential curve of luminosity for the Dymnikova, Fan-Wang and Schwarzschild BHs.

Now we compare the flux from the accretion disk of RBHs with the one of a Schwarzschild BH (Fig. \ref{fig:FluxT} (left panel)). The accretion disk around RBHs emits more energy than the Schwarzschild BH. This occurs as a result of the RBH's inner edge being moved toward smaller values of $x$ in contrast to the Schwarzschild BH.

The temperature of accretion disks in the unit of disk accretion rate is displayed in Fig \ref{fig:FluxT} (right panel). Here we compare the temperature curve of the accretion disk around the Schwarzschild BH (black solid curve) with the one around the Dymnikova and Fan-Wang BHs (blue and red solid curves). As seen in the figure, the temperature of the Fan-Wang BHs is greater than that of the Dymnikova and Schwarzschild BHs due to the shift of the inner edge of the accretion disk toward the smaller values of $x$. This is due to the ISCO of the Fan-Wang BH being less than that of the Dymnikova BH (See Tab \ref{tab:1}).

A comparison between the luminosity derivative of the classical BH and the one of the Dymnikova and Fan-Wang BHs is shown in Fig. ~\ref{fig:Diflum}. As seen in Fig.~\ref{fig:Diflum} the derivative of the luminosity of the accretion disk of the Dymnikova and Fan-Wang BHs differs from that of a Schwarzschild BH.  The luminosity derivative of the Dymnikova and Fan-Wang BHs is higher than the Schwarzschild BH.\\

\section{Efficiency of Mass Conversion to Radiation}\label{sez4}

We compute the efficiency of converting mass to radiation, symbolized as $\epsilon$, by analyzing the energy dissipation of a test particle moving from infinity to the disk's inner edge. It is assumed that all photons emitted from the disk's surface can reach infinity. Thus, the particle's energy tends to unity, or $E_\infty\approx1$, as the radius gets closer to infinity. So, we write

\begin{eqnarray}
\label{eq.028}
\epsilon=\frac{E_\infty-E_{isco}}{E_\infty}\approx1-E_{isco},
\end{eqnarray}

For the critical values of Dymnikova and Fan-Wang's free parameters, we calculate the appropriate bounds on specific energy in ISCOs, $E_{isco}$, and the conversion efficiency of the accreting mass into the corresponding radiation, $\epsilon$, see Tab. \ref{tab:1}.

This trend seems to be the opposite of what the Schwarzschild BH predicted, and it is especially noticeable for the Dymnikova and Fan-Wang BHs. 
 
\begin{table}[ht]
\begin{center}
\caption{Specific energy in the innermost stable circular orbit $E_{isco}$ and mass-to-radiation conversion efficiency $\epsilon$ for the Dymnikova and Fan-Wang RBHs with different values of free parameters, and comparison with the Schwarzschild-space-time.}
\vspace{3 mm}
\label{tab:1}
\begin{tabular}{|c|c|c|c|}
\hline
  \hline			
  $Free\,\,Parameter$ & $x_{isco}$ & $E_{isco}$ & $\epsilon(\%)$ \\
  \hline
  \hline
  $\bar{l}_{D}=0.45$ & 3.60  & 0.9112 & 8.8767 \\
  $\bar{l}_{D}=0.35$ & 4.54 & 0.9238 & 7.6209 \\
  $\bar{l}_{D}=0.26$ & 4.85 & 0.9308 & 6.9197 \\
  $\bar{l}_{D}=0.17$ & 5.29 & 0.9359 & 6.4081 \\  \hline
  $\bar{l}_{FW}=8/27$ & 2.67 & 0.8856 & 11.4314 \\
  $\bar{l}_{FW}=6/27$ & 3.74 & 0.9130 & 8.6982 \\
  $\bar{l}_{FW}=4/27$ & 4.57 & 0.9270 & 7.3036 \\
  $\bar{l}_{FW}=2/27$ & 5.31 & 0.9361 & 6.3874 \\ \hline
    $\bar{l}_{D},\bar{l}_{FW}=0$& $6$ & 0.9428 & 5.7191 \\
  \hline
\hline
  \end{tabular}
  \end{center}
\end{table}

Furthermore, increases in flux, temperature, and differential luminosity, as well as the conversion efficiency of accreting mass to radiation, are directly related to the displacement of the ISCO to lower values.

It turns out that the key quantity for describing the physical characteristics (both optical and structural) of an accretion disk is the $r_{ISCO}$, which is, in turn, defined by the parameters of the gravitational source. In the case of the Dymnikova and Fan-Wang solutions, in addition to the mass of the source, additional length parameters related to electric and magnetic charges emerging from NLED substantially alter the geometry around RBHs in a similar manner to an electric charge in the Reissner--Nordstr\"{o}m  BH or color charges in the charged dilatonic BH solutions \cite{2011PhRvD..83b4021P,2024EPJC...84...19B}. Hence, any BH with any form of charge induces additional gravity.


\section{Final outlooks and perspectives}\label{sez5}

As a consequence of the modern impressive interest around RBHs, we here explored the thermodynamic properties of thin accretion disks around precise regular spacetimes that can be obtained from NLED Lagrangian scenarios and may turn to be interesting in modelling compact objects without singularities. 

Recently, the study of accretion disks surrounding RBHs has assumed considerable importance in astrophysics, as this research avenue provides chances to deepen our understanding of RBHs as distinct entities from conventional BHs.

As regular solutions, we here considered the Fan-Wang and Dymnikova RBHs, evaluating for them the simplest accretion disk model proposed by Novikov-Thorne and Page-Thorne. 

In so doing, we calculated temperature, differential luminosity, and mass conversion efficiency into radiation to compare the actual findings with previous literature. 

Our findings indicated that introducing non-zero free parameters, distinct for Fan-Wang and Dymnikova metrics and here denoted as ${l}_{D}$ and ${l}_{FW}$, resulted in a reduction in the corresponding ISCOs. Consequently, we interpreted that non linear couplings of gravity with electrodynamics can affect the measurements of the expected overall spectral properties.

Nevertheless, we uncovered another noteworthy result due to the non-zero free parameters, ${l}_{D}$ and ${l}_{FW}$, namely, an increase in energy emitted from the disk surface around the RBH solution. 

Remarkably, utilizing specific values for the free parameters, and precisely ${l}_{D}=0.452$ and ${l}_{FW}=8/27$, we found an increase in the temperature of the accretion disk surrounding RBHs. Additionally, incorporating non-zero values for these parameters resulted in an enhancement in the differential luminosity of the accretion disk around Dymnikova and Fan-Wang BHs.

Additionally, we observed that non-zero values of the free parameters increased the efficiency of mass conversion into radiation compared to the Schwarzschild spacetime. 

Thus, the here-involved particular solutions obtained from NLED may be seen to be more efficient in converting mass into radiation than Schwarzschild BHs, suggesting how future explorations can check whether a given  compact object can be described by such solutions or not. 

In view of our findings, future investigations will go through  diverse scenarios involving rotating RBHs, encompassing the impact of magnetic fields and dark matter on the physical characteristics of the accretion disk. Furthermore, embracing alternative and potentially more intricate accretion disk models will be crucial to probing potential deviations from our current approach. By scrutinizing these alternative models, we aim to elucidate differences arising as we move away from the assumption of spherical symmetry within the employed spacetimes. This proposal  will clearly shed light on the expected complexities of accretion disk dynamics, providing valuable insights into the behavior of RBH accretion disk systems.

\begin{acknowledgments}
YeK acknowledges Grant No. AP19575366, TK acknowledges the Grant No. AP19174979, and GS acknowledges Grant No. AP19680128, and KB and AU acknowledge the Grant No. BR21881941 from the Science Committee of the Ministry of Science and Higher Education of the Republic of Kazakhstan.
\end{acknowledgments}

\bibliography{0biblio}

\end{document}